\documentclass{article}
\usepackage[cp1250]{inputenc}
\usepackage[T1]{fontenc}
\newcommand{\bd}{\begin{displaymath}}
\newcommand{\ed}{\end{displaymath}}
\newcommand{\ee}{\end{equation}}
\newcommand{\be}{\begin{equation}}
\newcommand{\laa}{\lambda_1}
\newcommand{\lb}{\lambda_2}
\newcommand{\lc}{\lambda_3}
\newcommand{\ld}{\lambda_4}
\newcommand{\lp}{\lambda_5}
\newcommand{\lcz}{\lambda_{45}}
\newcommand{\lczp}{\lambda_{345}}

\newcommand{\bea}{\begin{eqnarray}}
\newcommand{\eea}{\end{eqnarray}}
\newcommand{\bes}{\begin{subequations}}
\newcommand{\ees}{\end{subequations}}
\newcommand{\bear}{\begin{equation}\begin{array}}
\newcommand{\eear}[1]{\end{array}\label{#1}\end{equation}}
\newcommand{\fr}[2]{\dfrac{{ #1}}{{ #2}}}

\newcommand{\fn}[1]{\footnote{{\sf #1}}}
\usepackage{amsmath}
\usepackage[lofdepth,lotdepth]{subfig}
\usepackage{authblk}

\usepackage{epsfig}

\begin{document}
\title{New results for the Inert Doublet Model%
\thanks{Presented by M. Krawczyk at the XXXV International Conference of Theoretical Physics
"MATTER TO THE DEEPEST": Recent Developments in Physics of Fundamental Interactions, Ustroń'11
September 12-18, 2011}%
}
\author{Bogumiła Gorczyca, Maria Krawczyk}
\affil{\emph{Faculty of Physics, University of Warsaw}\\
\emph{Warsaw, Poland}}

\maketitle
\begin{abstract}
New results for the Inert Doublet Model (IDM) are discussed. It is  very special among the $D$-symmetric 2HDMs, offering a good DM candidate. New unitarity constraints were derived for the IDM and SM-like light Higgs boson scenario in the Mixed Model.
\end{abstract}

\section{Two Higgs Doublet Models}
Among the standard models of the elementary particle interactions the most popular are the one with one Higgs (scalar) doublet (The Standard Model SM = 1HDM) and with two such doublets  (2HDMs, including Minimal Supersymmetric Standard Model (MSSM)). In 2HDMs there are five scalars - three neutral and two charged ones. The lightest neutral scalar is often SM-like, what makes such models particularly interesting nowadays.

The Brout-Englert-Higgs mechanism describing spontaneous breaking of the  EW symmetry allows in 2HDM's for breaking of the $U(1)_{QED}$ symmetry, in contrast to the 1HDM. In these models  two scalar doublets of  SU(2), with weak hypercharge equal 1, can be involved in generating masses of the gauge bosons $W^\pm$ and  $Z$. Fermion masses are generated via Yukawa interactions in various ways,  leading to various models: Model I, II, III, IV,X,Y,.. depending on how the doublets couple to fermions. Typically, in order to avoid FCNC at the tree level, some discrete symmetries are imposed on a Lagrangian.
Here we will consider the  Lagrangian, which is symmetric under such $Z_2$  transformation,  where one of the scalar doublets changes sign, while all other fields (the other scalar doublet and all SM-fields) are unchanged. This allows us to consider a case of the Inert Doublet Model (IDM), in which such a $Z_2$ symmetry is respected not only at the Lagrangian level but also in the vacuum \cite{Ma78}. IDM  is unique among 2HDMs  as it predicts existence of a stable particle - a good candidate for the dark matter (DM) \cite{barbieri,Ma,Honorez,dolle,ds}.

We will call the scalar doublet which  changes sign under the transformation $\phi_D$,  while  the other scalar doublet we will denote as $\phi_S$. The corresponding $Z_2$-symmetry will be called the $D$-symmetry. The scalars will be universally  denoted by $h, H, A, H^\pm$ in all 2HDMs considered here.

We can consider the following $D$-symmetric potential \cite{nasza}:
\bes\label{baspot}
\bear{c}
V=-\fr{1}{2}\left[m_{11}^2(\phi_S^\dagger\phi_S)\!+\! m_{22}^2(\phi_D^\dagger\phi_D)\right]+\\[2mm]
\!\!\!+\fr{\lambda_1}{2}(\phi_S^\dagger\phi_S)^2\!+\!\fr{\lambda_2}{2}(\phi_D^\dagger\phi_D)^2+\!
\lambda_3(\phi_S^\dagger\phi_S)(\phi_D^\dagger\phi_D)\!+\!\\[2mm]
\!+\!\lambda_4(\phi_S^\dagger\phi_D)(\phi_D^\dagger\phi_S) +\fr{\lambda_5}{2}\left[(\phi_S^\dagger\phi_D)^2\!
+\!(\phi_D^\dagger\phi_S)^2\right],
\eear{baspot1}
with all  parameters real  and with an additional condition
$
\lambda_5<0\,.\label{baspot2}$
\ees
The IDM is realized in some regions of parameter space of this potential.
We will consider also other  possible vacuum states of such potential, realized at another values of parameters. This allows to consider  possible temperature evolutions of vacua and transitions between them, see below and in \cite{nasza,ds}.

\section{Extrema and vacua}
Extrema of the 2HDM potential with an explicit $D$-symmetry can be found as usual: first, one
finds extrema, then minima and then the global minimum, which is the vacuum. Positivity (stability) constraints on the $V$ are as follows
 \begin{eqnarray}
& \lambda_1>0\,, \lambda_2>0, \, R + 1 >0, \,\lambda_{345}=\lambda_3+\lambda_4+\lambda_5,\, R = \lambda_{345}/\sqrt{\lambda_1 \lambda_2}. &
\end{eqnarray}
The \emph{extremum} respecting these constraints, which has the lowest energy, is the vacuum of the system \cite{nasza}.

In general the extrema have the following form:
\bear{c}
        \langle\Phi_S\rangle =\dfrac{1}{\sqrt{2}}\left(\begin{array}{c} 0\\
        v_S\end{array}\right),\quad \langle\Phi_D\rangle
        =\dfrac{1}{\sqrt{2}}\left(\begin{array}{c} u \\ v_D
        \end{array}\right),
\eear{genvac}
with $v_S >0$ and $v^2=v_S^2+|v_D^2|+u^2$, $v$=246 GeV. Properties of   extrema  respecting  and violating $U(1)_{QED}$ symmetry  ($u=0\,{\rm{and}} \,u\neq 0$, respectively) are presented in Table \ref{vacua}.

\begin{table}
\begin{center}
{\renewcommand{\arraystretch}{1.5}
\begin{tabular}{|c|c|p{7.5cm}|}
\hline
\hline
\multicolumn{2}{|c|}{Extrema} \\ \hline \hline
\textit{EW symmetric}: $EW\! s$ & $u=0,\quad v_D=0, \quad v_S=0$  \\ \hline
 \textit{Inert}: $I_1$ & $u=0,\quad v_D=0,\quad v_S^2=\fr{m_{11}^2}{\lambda_1}$ \\ \hline
\textit{Inert-like}: $I_2$ & $u=0,\quad v_S=0,\quad v_D^2=\fr{m_{22}^2}{\lambda_2}$  \\ \hline
\textit{Mixed}: $M$ & $u=0,\quad v_S^2, v_D^2 > 0 $  \\ \hline
\textit{Charged}: $Ch$ & $ u, v_S^2 > 0$  \\ \hline
\hline
\end{tabular}}
\caption{General properties of extrema, following \cite{nasza}. \label{vacua}}
\end{center}
\end{table}

\begin{figure}
\centering
\includegraphics[width=0.5\textwidth,angle=-90]{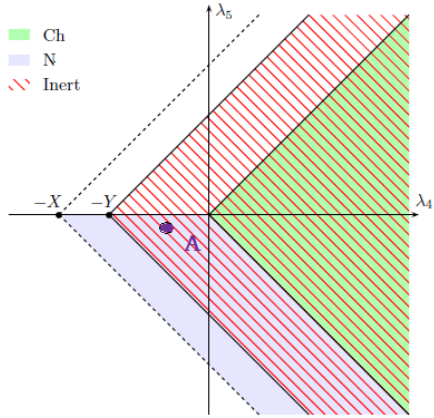}
\caption{Region allowed by the positivity constraints (on the right of the dotted lines). Allowed regions for extrema:  the Inert (Inert-like) (hatched area), Mixed (shaded area for $\lambda_4+\lambda_5<0$) and Charged (hatched shaded area for $\lambda_4\pm\lambda_5>0$). Point A corresponds to a possible today's Universe state.  \label{l45}}
\end{figure}

It is very useful to represent  extrema in the ($\lambda_4 , \lambda_5$) plane (Fig. \ref{l45}) \cite{MK-DS}.
In this figure positivity constrains lead to the bounds: $\lambda_4 ± \lambda_5 > - X$, where  $X=\sqrt{\lambda_1\lambda_2}+\lambda_3>0$; for the Inert vacuum
$Y = 2M_{H+}|_{Inert}^2/v^2>0$, while for the Charged one $\lambda_4 ± \lambda_5 >0 $. \fn{Here we show $\lambda_5 >0$ regions allowed for Inert(Inert-like) and Charged vacua, which are symmetric to the $\lambda_5 <0$ ones, with a change of roles of $H$ and $A$ particles.}
Note the overlap of the regions where the Inert (or Inert-like) vacuum can be realized   with the corresponding ones allowing for Mixed  and/or Charged  vacua.

If Nature is described today by the  $D$-symmetric 2HDM Lagrangian (with                                          Model I  of the Yukawa interactions, where only the $\phi_S$ couples  to fermions) then
the question is, which vacuum (phase) is realized today? Definitely charge breaking phase is not
a good candidate, as  here photon would be massive and electric charge would not be conserved. Among
  neutral phases only the Inert one, being in agreement with accelerator and astrophysical data, offers a good neutral DM candidate (for $\lambda_5<0$ it is a $H$).
  Inert-like phase is excluded as here all the fermions would be massless, on the other hand Mixed phase is in agreement with the accelerator  data.

We have  considered  evolution of the Universe in 2HDM, during its cooling down  from the EW symmetric phase to the present Inert phase \cite{nasza}, see also \cite{iv2008}.
For this purpose thermal evolution of the explictly $D$-symmetric Lagrangian was considered
in the simplest approximation, where  only mass terms in V vary with temperature like $T^2$, while parameters $\lambda's$ are fixed. In Fig.\ref{l45} we show a possible position of the today's Universe.
In the past it could go through various phases in one, two or three phase transitions. We found that the first  phase transitions (i.e. from the EWs phase) are all of the 2nd order. One should, however, consider other thermal corrections beyond the $T^2$ approximation; preliminary results were obtained recently \cite{GG}, suggesting that the type of these transitions may change.

\section{Inert Doublet Model}
In the IDM  $D$-symmetry is conserved and one can assign $D$-parity to all particles.
The $\phi_S$, with $D$-parity even,  plays a role of Higgs doublet in the SM, with one Higgs (SM-like) particle $h$  (with  $M_{h}^2=\lambda_1v^2= m_{11}^2$).
The second doublet $\phi_D$, with zero vacuum expectation value,  is $D$-odd, it contains 4 scalars (not Higgs particles!) and the lightest particle among them is stable and can be a dark matter particle. We call all these scalars - dark scalars; their masses are given by
\be
M_{H^\pm}^2=\fr{\lambda_3 v^2-m_{22}^2}{2},\,
M_{A}^2=M_{H^\pm}^2+\fr{\lambda_4-\lambda_5}{2}v^2\,, M_{H}^2=
M_{H^\pm}^2+\fr{\lambda_4+\lambda_5}{2}v^2\,.
\ee\label{massesA}
Couplings among scalars are given by $\lambda's$: $\lambda_1$ is proportional to  $hhh$ coupling and fixed by the mass of $h$, $\lambda_{345}$ describes trilinear couplings of $h$ with dark scalars,  while $\lambda_2$ appears only in quartic selfcouplings of the dark scalars.

Theoretical constraints of the IDM  arising from the positivity (stability) condition and conditions for the Inert vacuum were discussed above and can be found in \cite{nasza,ds}. The important new unitarity constraints were obtained in \cite{BG} and are presented below.  Here we would like to mention agreement of this model with precision EW data, in form of $S,T,U$, see eg. \cite{barbieri,Ma,dolle}.

Phenomenologically, testing the IDM at present and future colliders can be performed by
precise measurements of properties of the SM-like $h$ and by direct search of dark scalars' pairs.
There exist some constraints from LEP II (masses H versus A) \cite{LEPII-IDM}, as well as analysis on DM \cite{Honorez,dolle,ds}.

\section{Unitarity constraints for the IDM  and the Mixed Model}
Here we present new results on unitarity constraints for the $D$-symmetric 2HDM potential \cite{BG}. It updates the previous analyses for parameters $\lambda$ \cite{huffel}. We have applied the standard approach \cite{lqt}, using equivalence theorem to deal with Goldstone bosons instead of longitudinal gauge bosons and neglecting the trilinear couplings. We applied, for the first time, unitarity constraints for the IDM and for the SM-like scenario within the Mixed Model (based on the Mixed vacuum and the Model II of Yukawa interactions).

\begin{figure}
\centering
\subfloat[Allowed masses of $H^{\pm}$ and $A$ in the IDM with $m_{22}^2=0$ (dark points) and $m_{22}^2=-10^6\,\textrm{GeV}^2$ (pale points).\label{inert}]{
\includegraphics[width=0.45\textwidth]{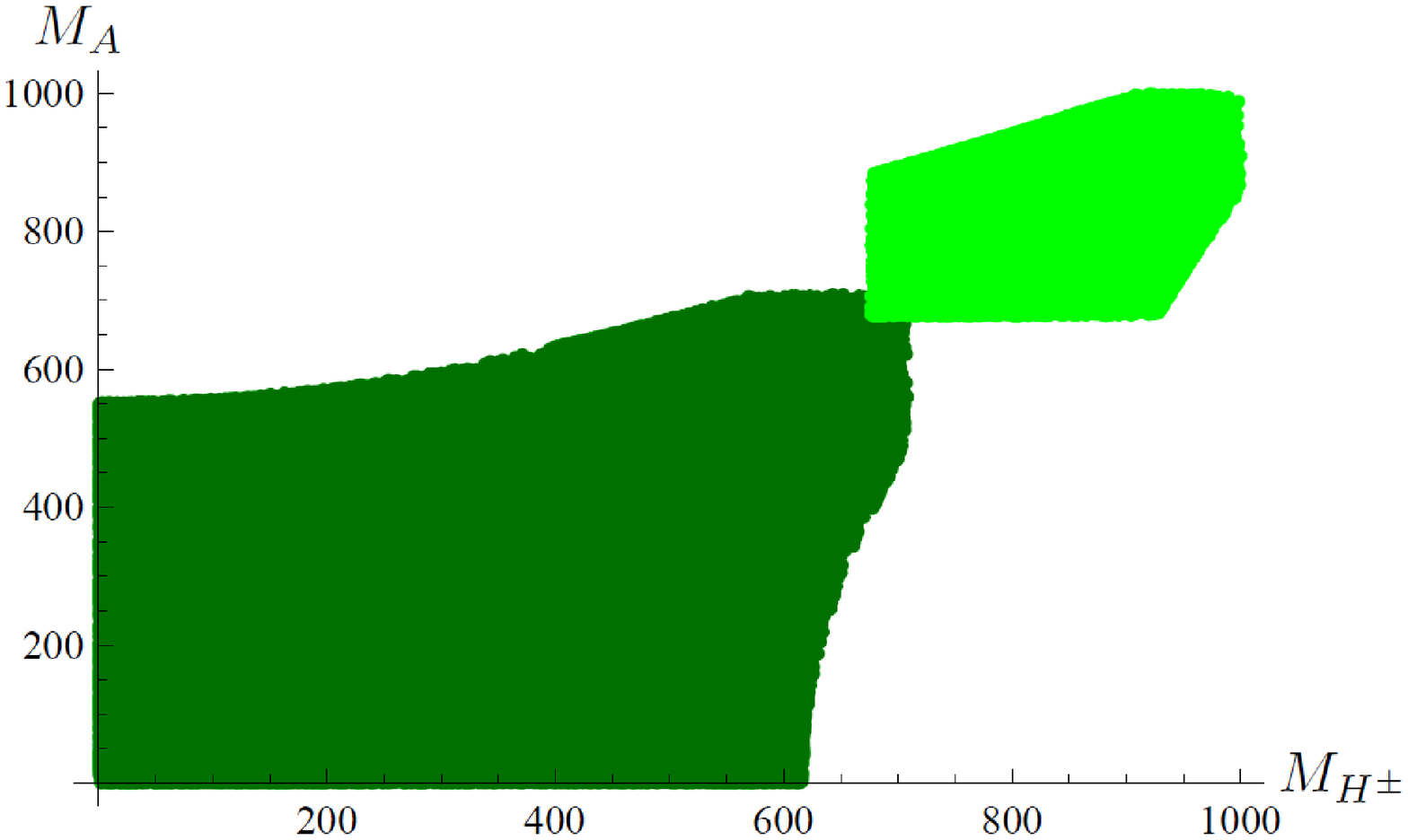}}\quad
\subfloat[Allowed masses of $H^{\pm}$ and $H$ in the Mixed Model.\label{mixed}]{
\includegraphics[width=0.45\textwidth]{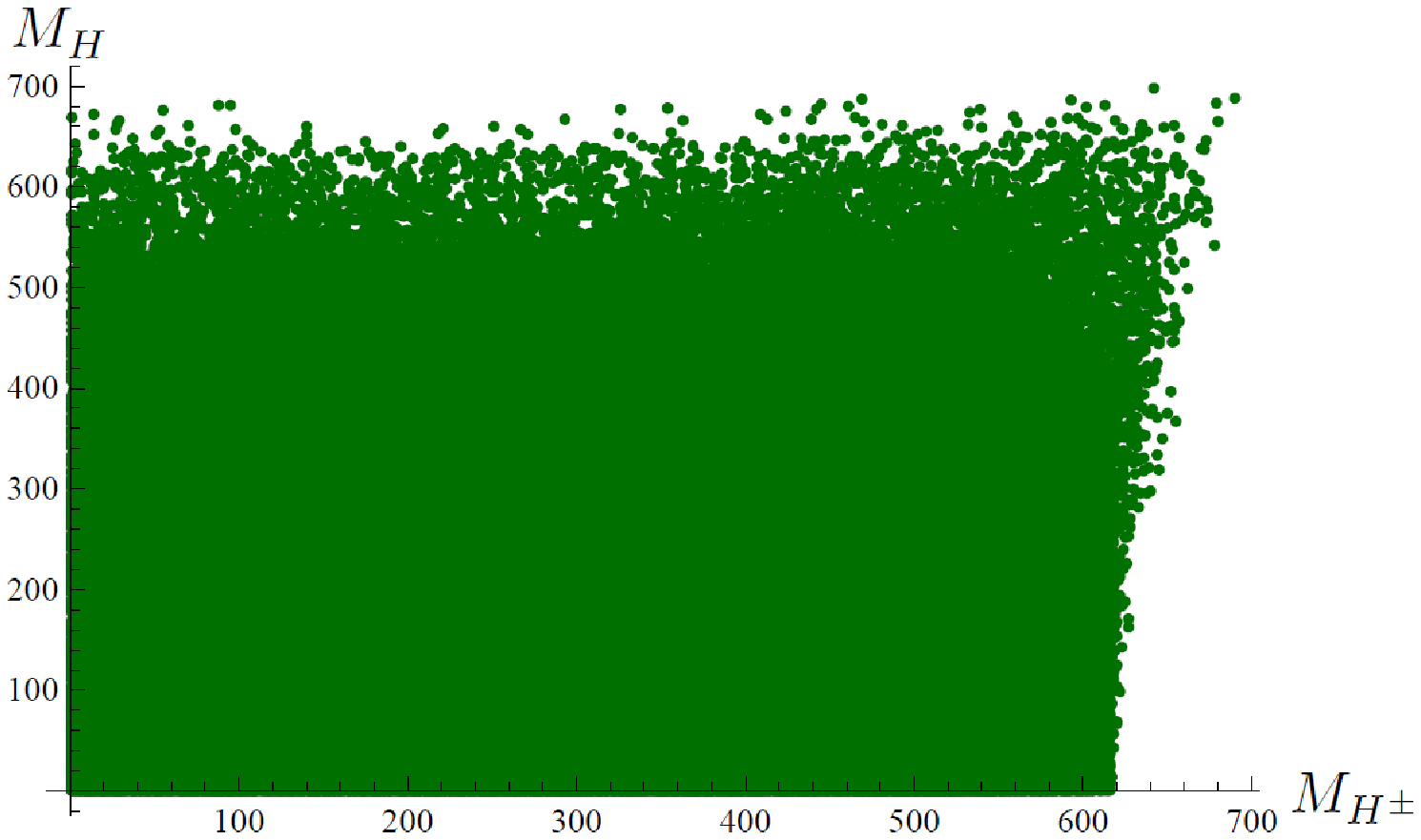}}
\caption{Regions of masses allowed by the unitarity constraint for  the IDM (left) and the Mixed Model (right). \label{ploty}}
\end{figure}

Full tree-level high energy scattering matrix (of dimension 25) for the scalars was considered,
including the double charged initial/final states not studied previously. Diagonalization thereof leads to 12 distinct eigenvalues being functions of the quartic couplings (or equivalently of the parameters $\lambda$) ~\cite{kanemura, akeroyd}.  Applying the standard unitarity condition $|\Re( a^{(j)}(s))|\leq\frac{1}{2}$ to these eigenvalues yields a set of inequalities for $\lambda$'s or, if different set of parameters is chosen, for the masses of scalar particles. These inequalities were solved numerically, probing statistically a large range of values of the parameters (as in \cite{akeroyd}) and taking into account  the vacuum stability conditions and conditions determining the type of vacuum, as discussed in Sec. 3.
The results of the scans give bounds on the values of the $\lambda$ parameters and masses and correlations between them (see Fig.\,\ref{ploty}), not considered in previous analyses \cite{kanemura, akeroyd}.
The constraints obtained for the $\lambda$'s are more stringent  then the previous ones and read:
\be
\begin{array}{r@{\;\leq\;}c@{\;\leq\;}l}
0&\laa&8.38,\\
0&\lb&8.38,\\
-5.85&\lc&16.33,\\
-15.82&\ld&5.93,\\
-8.21&\lp&0.\\
\end{array}
\ee
Similarly, the following combinations of $\lambda$'s are constrained 
($\lambda_{ij}=\lambda_i +\lambda_j$)
\be\label{lami}
\begin{array}{r@{\;\leq\;}c@{\;\leq\;}l}
-7.90&\lczp&11.31,\\
-16.37&\lcz&0,\\
-7.45&\lambda_{34}&12.55,\\
\end{array}
\ee
which in the IDM  correspond directly to the bounds on quartic couplings between physical fields
\fn{ $\lczp$ represents a coupling of $hhHH$, $AAGG$, 
$\lcz$ is a coupling of a vertex containing $H^+G^-$ or $G^+H^-$ and $\lambda_{34}$ is a coupling of $G^+G^-H^+H^-$}.

In the IDM the bounds on masses of the scalars depend on one additional  parameter $m_{22}^2$ (Eq.\,(\ref{massesA})). However, this dependence is negligible for $|m_{22}^2|\leq 10^4\,\textrm{GeV}^2$. The upper bounds on masses of $H^{\pm}$ and $A$ for this case (when $M_h=120\,\textrm{GeV}$ or $M_h\in[114,\,145]\,\textrm{GeV}$) are of order of $700\,\textrm{GeV}$ and for $M_H$ of order of $600\,\textrm{GeV}$. This shows that in a wide region of values of $m_{22}^2$ the possibility of existence of a very heavy (with mass over $800\,\textrm{GeV}$) dark matter particle is excluded. The region of masses allowed by the unitarity condition for the cases with $m_{22}^2=0$ and $m_{22}^2=-10^6\,\textrm{GeV}$ is shown in Fig.\,\ref{inert}.

In the Mixed Model the upper bounds for the heavy scalars' ($H^{\pm}$, $H$, $A$) masses are of order of $700\,\textrm{GeV}$ and for the $h$ boson of order of $500\,\textrm{GeV}$ (as in \cite{nowa}). The region of masses of $H^{\pm}$ and $H$ allowed by the unitarity condition is shown in Fig.\,\ref{mixed}.

In addition, we consider the SM-like scenario of the Mixed Model, with the condition $\sin(\beta-\alpha)=1$. Then the $h$ boson couples (at the tree-level) to fermions and gauge bosons exactly as  the SM Higgs boson and  the experimental constraints for the SM Higgs mass can be applied to $h$: $M_h\in[114,\,145]\,\textrm{GeV}$. Unitarity constraints lead to  the upper bounds of about   $ 600\,\textrm{GeV}$ for $M_H$ and $M_H^{\pm}$, which are lowered as compared to the arbitrary $\sin(\beta-\alpha)$ case, and do not bound $M_h$ any further. 

\section{Conclusions}
A significant progress has been obtained recently in understanding the underlying structure of the simple extensions of the SM with two scalar doublets. IDM is very special among the $D$-symmetric 2HDMs, offering a good DM candidate. New unitarity constraints were derived for the IDM and SM-like
light Higgs boson scenario in the Mixed Model. \\

We are grateful to  D.~Sokołowska, P.~Chankowski, G.~Gil, I.~F.~Ginzburg, K.~A.~Kanishev and H.~Haber for fruitful discussions. MK thanks organizers for this excellent conference.
Work was partly supported by Polish Ministry of Science and Higher Education Grant N N202 230337.\\

\textit{
This paper had originally been  published in Acta Physica Polonica (Acta Phys. Pol. B 42 (2011) 2229). However, we found an error in our calculations, which modified the results for the Mixed Model. The corrections were pulished in Erratum (Acta Phys. Pol. B 43 (2012) 481). In the present article corresponding erros have been removed.}

\end{document}